\documentclass[prl, superscriptaddress, showpacs, floatfix, twocolumn]{revtex4-2}
\usepackage[utf8]{inputenc}
\usepackage{graphicx}
\usepackage{hyperref}
\usepackage{amsmath}
\usepackage{amssymb}
\usepackage{tabularx}
\usepackage{booktabs}

\newcommand{\ols}[1]{\mskip.5\thinmuskip\overline{\mskip-.5\thinmuskip {#1} \mskip-.5\thinmuskip}\mskip.5\thinmuskip}
\newcommand{\olsi}[1]{\,\overline{\!{#1}}}

\hypersetup{
    colorlinks=true,
    linkcolor=blue,
    filecolor=blue,
    urlcolor=blue,
    citecolor=blue,
}

\begin{document}
\title{Fragile topologically flat band in the checkerboard antiferromagnetic monolayer FeSe}
\author{Aiyun Luo}
\affiliation{Wuhan National High Magnetic Field Center $\&$ School of Physics, Huazhong University of Science and Technology, Wuhan 430074, China}
\author{Zhida Song}
\affiliation{Department of Physics, Princeton University, Princeton, New Jersey 08544, USA}
\author{Gang Xu}
\affiliation{Wuhan National High Magnetic Field Center $\&$ School of Physics, Huazhong University of Science and Technology, Wuhan 430074, China}

\begin{abstract}
By means of the first-principles calculations and magnetic topological quantum chemistry,
we demonstrate that the low energy physics in the checkerboard antiferromagnetic (AFM) monolayer FeSe,
very close to an AFM topological insulator that hosts robust edge states,
can be well captured by a double-degenerate fragile topologically flat band just below the Fermi level.
The Wilson loop calculations identify that such fragile topology is protected by the $S_{4z}$ symmetry,
which gives rise to an AFM higher-order topological insulator that support the bound state with fractional charge $e/2$ at the sample corner.
This is the first reported $S_{4z}$-protected fragile topological material,
which provides a new platform to study the intriguing properties of magnetic fragile topological electronic states.
Previous observations of the edge states and bound states in checkerboard AFM monolayer FeSe can also be well understood in our work.
\end{abstract}

\maketitle

\textit{Introduction} ---
Topological matters have attracted extensive interest for their novel and robust bulk-boundary correspondence~\cite{PhysRevLett.95.226801,Bernevig1757,Konig766,zhang2009topological,Chen178}.
In particular, the interplay between the crystal symmetry and electronic bands gives rise to a variety of topological states,
such as the topological crystalline insulator~\cite{PhysRevLett.106.106802,hsieh2012topological,wang2016hourglass} and higher-order topological insulator (HOTI)~\cite{Benalcazar61,PhysRevB.96.245115,Schindlereaat0346,song2017,PhysRevLett.122.256402,wieder2018,PhysRevB.100.235137}.
Moreover, when the magnetism and topological bands are entangled together,
it will greatly enrich more exotic topological states~\cite{PhysRevB.81.245209,Yu61,Chang167,li2010dynamical,Watanabe2018,xu2020high,elcoro2020magnetic},
such as the quantum anomalous Hall effect~\cite{Chang167,Deng2020},
axion topological insulators~\cite{Zhang2019,Liu2020}, and magnetic topological semimetals~\cite{Xu2011,Zou2019}.
More recently, a new type of topological state, namely the fragile topology~\cite{PhysRevLett.121.126402,Cano2018}, has been proposed.
Different from the stable topology, the fragile topology generally cannot exhibit robust edge states.
Meanwhile, it hosts the Wannier obstruction,
which means that it should undergo a gap closing to evolve into a trivial insulator.
The fragile topology shows many intriguing properties \cite{Song794,Peri797,Xie2020,Peri2021,Lian2020,Jonah2020}.
For example, the fragile topologically flat band contributes a nontrivial superfluid weight
in 2D superconductor~\cite{Xie2020,Peri2021}, and hence enhances the superconducting transition temperature.
Under a varying magnetic field, the corner states of the fragile Hofstadter topological system can pump into bulk by magnetic flux~\cite{Jonah2020}.
Besides the conceptual breakthrough, searching for more natural fragile topological materials is highly desirable.

As a wonderful platform to study the emergent phenomena among superconductivity, magnetism and topological bands,
FeSe has generated much attentions recent years, in which high-$T_c$ superconductivity~\cite{tan2013interface,he2013phase,ge2015superconductivity}, topological superconductivity~\cite{Wang2015,Xu2016,Zhang2018}, topological states~\cite{Hao2014,Zhang2019B,wang2016topological,liubin2018} have been reported.
In particular, Wang \emph{et al.} have reported that the monolayer FeSe in the checkerboard antiferromagnetic (cb-AFM) phase
is a 2D AFM topological insulator (TI) ~\cite{wang2016topological}.
After that, edge states are observed at the nematic domain boundaries of monolayer FeSe~\cite{liubin2018}.
A bound state near Fermi level is further observed at the end of boundaries.
However, the topological origin of the bound state has not been discussed,
and the topological nature in the cb-AFM monolayer FeSe remains elusive until now.

\begin{figure}
  \includegraphics[width=\columnwidth]{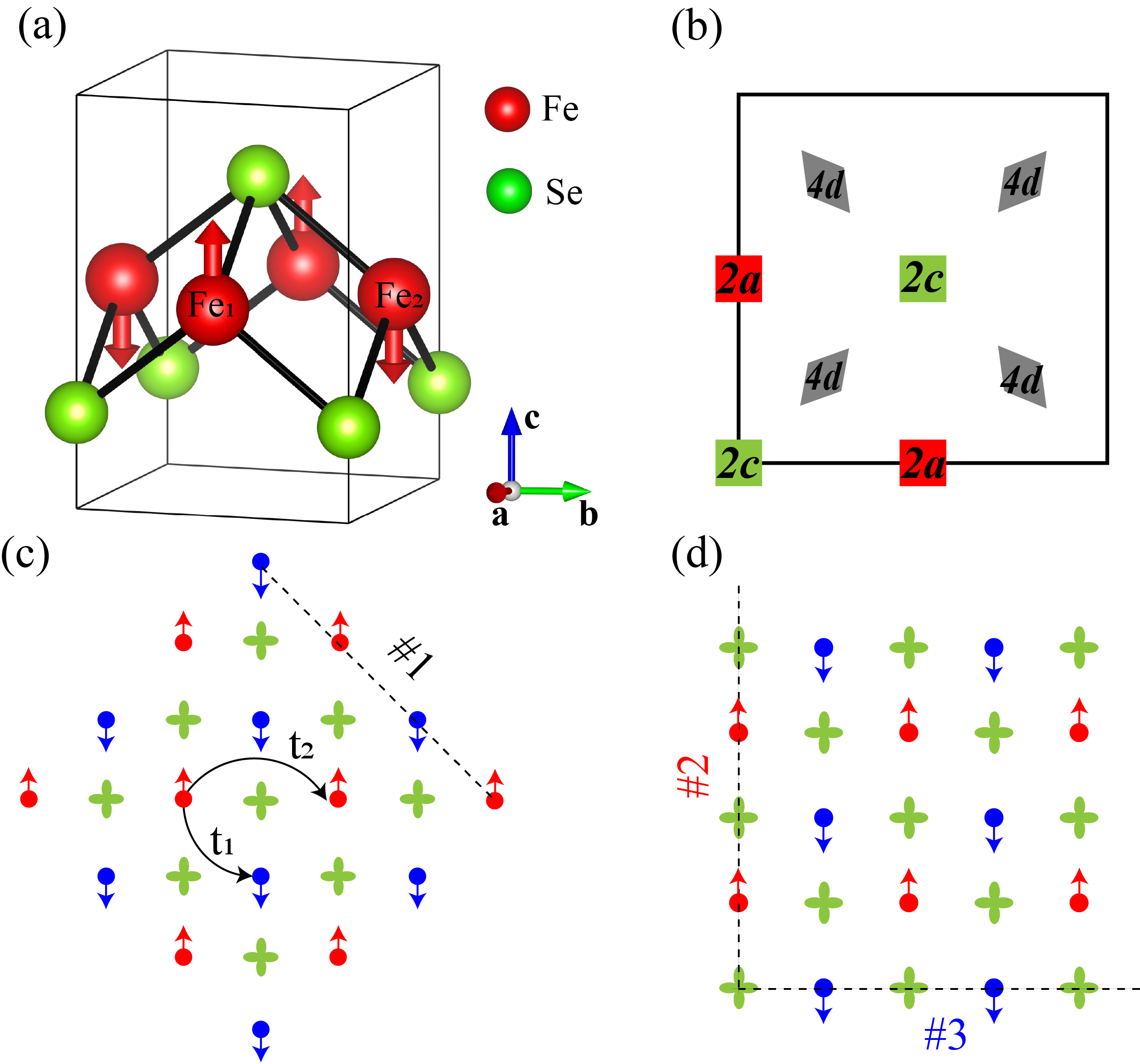}
  \caption{(a) The crystal and magnetic structure of monolayer FeSe. (b) Maximal symmetry Wyckoff positions in magnetic space group $P4'$/$n'm'm$.
  Since monolayer FeSe is 2D, only the $z=0$ Wyckoff positions are manifested.
  (c)$-$(d) Three distinct boundary cuts in FeSe. Cut \#1 represents the boundary of AFM edge. Cuts \#2 and \#3 represent the boundary of FM edge with
   spin-up and spin-down polarization along c-direction.}
  \label{fig:crystal}
\end{figure}

In this work, by using the first-principles calculations and magnetic topological quantum chemistry~\cite{Watanabe2018,xu2020high,elcoro2020magnetic}, we systematically study the topological properties of the cb-AFM monolayer FeSe. We demonstrate that their low energy physics can be well captured by a double-degenerate fragile topologically flat band just below the Fermi level, which gives rise to an AFM HOTI phase that support the bound state with fractional charge $e/2$ at the sample corner. The winding of square Wilson loop confirms that such fragile topology is protected by the $S_{4z}$ symmetry. This makes cb-AFM monolayer FeSe the first reported $S_{4z}$ protected fragile topological material, which may lead to intriguing responds to disorder and magnetic field. Further analysis find that cb-AFM monolayer FeSe is very close to a well-defined AFM TI with a nontrivial $S_z$ protected spin Chern number. These results can well explain the previous observations of topological edge states and bound state near the Fermi level~\cite{wang2016topological,liubin2018}, and also provide a new platform to study the intriguing properties of the magnetic fragile topology.

\textit{Crystal structure and computational method} ---
As illustrated in Fig.~\ref{fig:crystal}a and~\ref{fig:crystal}b,
both bulk and monolayer FeSe adopt the tetragonal lattice structure with the inversion symmetric space group $P4$/$nmm$ (No.~129),
in which Fe atoms occupy the Wyckoff position $2a$ (0, 0.5, 0), and Se atoms occupy the Wyckoff position $2c'$ (0, 0, $z$).
In this work, we mainly focus on the band structures and topological properties in the cb-AFM phase of monolayer FeSe as shown in Figs.~\ref{fig:crystal}a,c-d.
In this magnetic configuration, the nearest neighboring Fe ions are antiferromagnetically coupled to each other and the inversion symmetry is broken,
leading to a magnetic space group $P4'$/$n'm'm$ (No.~129.416~\cite{gallego2012magnetic}).
$P4'$/$n'm'm$ is generated from a space-time inversion $\mathcal{PT}$, a twofold screw $C_{2x}$ ($\{\hat{C}_{2x}|\frac{1}{2},0,0\}$),
and an improper fourfold rotation $S_{4z}$ ($\{\hat{C}_{4z}\hat{M}_z|0,\frac{1}{2},0\}$).
Our first-principles calculations are performed by the Vienna \emph{ab initio} simulation package (VASP) \cite{kresse1993ab, kresse1996efficient}.
Similar as in Ref.~\cite{liubin2018},
Perdew-Burke-Ernzerhof (PBE) type of the generalized gradient approximation + Hubbard $U$ (GGA+$U$)~\cite{perdew1996generalized, anisimov1991band}
is used as the exchange-correlation potential with U = 1.0 eV and J = 0.2 eV for the Fe $3d$-orbitals.
The crystal parameters $a = 3.85$ \AA~ and $z$ = 0.255 are used in all calculations.
The cut-off energy for the wave function expansion is set to $400$~eV,
and a $9\times9\times1$ k-mesh in the first Brillouin zone (BZ) is used for self-consistent calculations.
The SOC is considered self-consistently.
We then construct the Wannier functions without performing maximally localized procedure for the Fe $3d$-orbitals using WANNIER90 \cite{mostofi2014updated}.

\begin{figure}[t]
  \includegraphics[width=\columnwidth]{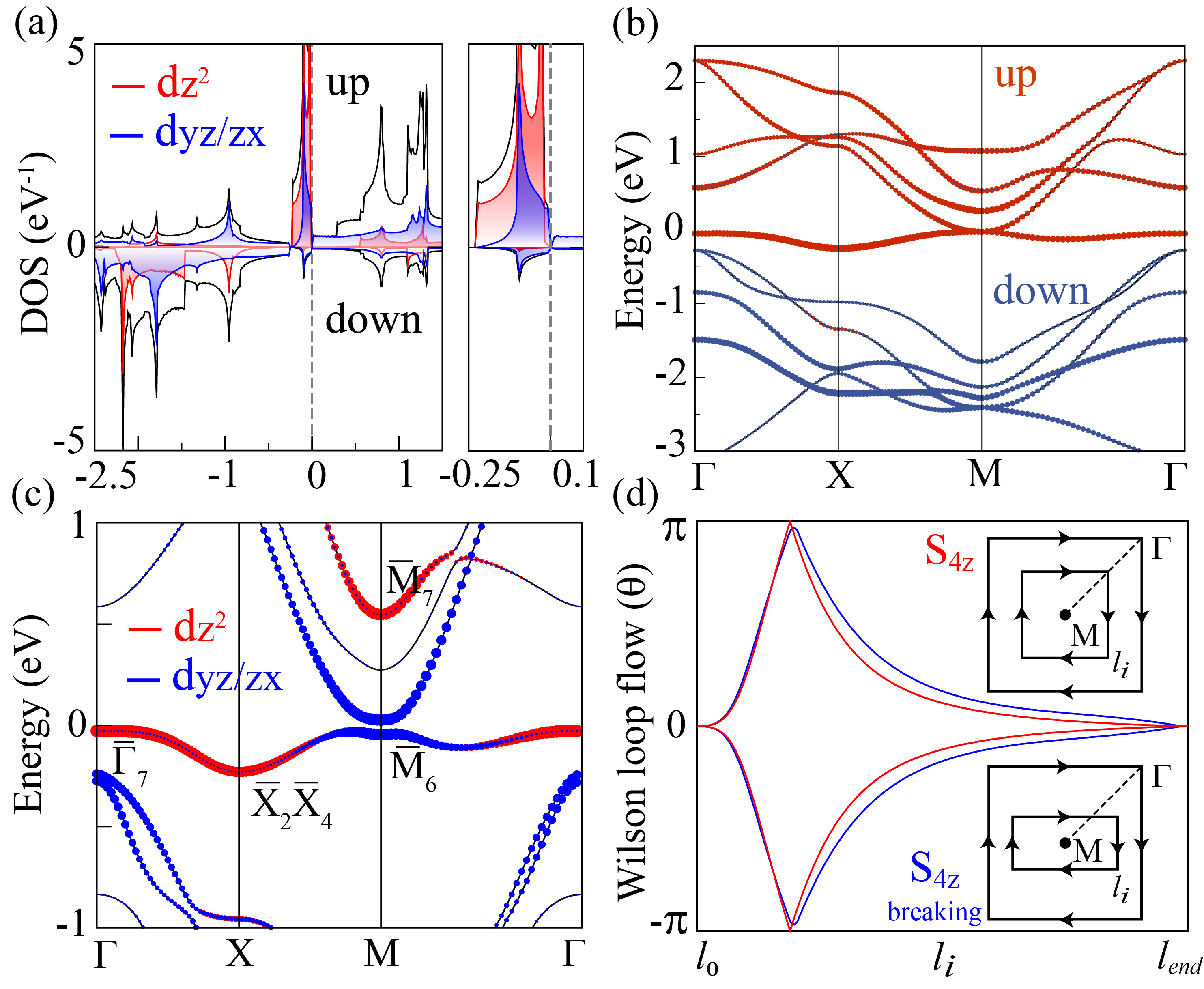}
  \caption{The electronic structures of cb-AFM monolayer FeSe.
  (a) Projected DOS of the $3d$-orbitals on Fe$_2$ for non-SOC calculations.
  (b) Spin-polarized band structures with the red (blue) color indicating the minority (majority) spin states.
  (c) Band structures with SOC, the red and blue color indicate the weights of $d_{z^2}$-orbital and $d_{yz/zx}$-orbital, respectively.
  (d) The Wilson loop spectra of the isolated flat band. The red (blue) lines represent the square (rectangle) Wilson loop spectra with loops patch in
  the upper (lower) right inset panel, where the arrow indicate the direction of loops.}
  \label{fig:fragile}
\end{figure}

\textit{Electronic structures and fragile topology} ---
First, we carry out the non-SOC calculations of the cb-AFM monolayer FeSe,
and plot the corresponding projected density of states (DOS) of the $3d$-orbitals on Fe$_2$ ion (see Fig.~\ref{fig:crystal}a) in Fig.~\ref{fig:fragile}a.
It gives rise to about of 2.5~eV Zeeman splitting between the majority spin states and minority spin states.
As a result, most majority spin states are pushed blow $-0.25$ eV,
while the states between $-0.25 \sim 2.25$ eV are mainly contributed by the minority spin $3d$-orbitals.
Further analysis demonstrates that the low-energy bands around the Fermi level (0 eV) are dominated by the minority spin of $d_{z^2}$- and $d_{yz/xz}$-orbitals,
which have very weak hybridization with the majority spin components as shown in Fig.~\ref{fig:fragile}a.
These feature can also be represented by the spin-polarized band structures as shown in Fig.~\ref{fig:fragile}b,
which indicates that five majority spin bands are fully occupied and only one minority band is occupied.
Such occupancy corresponds to the $3d^6$ configuration of the Fe$^{2+}$ ion very well.
Further detailed orbital components analysis demonstrates that the one occupied minority band is mainly contributed by the $d_{z^2}$-orbital,
except a band inversion with the $d_{yz/xz}$-orbitals at the $M$ point near the Fermi level (see Fig.~\ref{fig:fragile}c).
The band inversion leading to a semimetallic feature with a parabolic touching at the $M$ point as shown in Fig.~\ref{fig:fragile}b.
When the SOC interaction is included, the gapless touching of $d_{yz}$- and $d_{xz}$-orbitals at $M$ point is opened
that arise an insulating gap about of $50$~meV.
Since the dispersion of the $d_{z^2}$ band is very weak,
the gap opening also leads to a very flat isolated band just locate below the Fermi level, as shown in Fig.~\ref{fig:fragile}c.
We further note that each band is double-degenerate in the presence of $\mathcal{PT}$ symmetry.

To investigate the topological properties of the cb-AFM monolayer FeSe,
we employ the symmetry analysis under the theory of magnetic topological quantum chemistry~\cite{Watanabe2018,xu2020high,elcoro2020magnetic},
in which the magnetic elementary band representations (MEBRs) as the basis of magnetic atomic insulators (AIs) are defined.
Within this theory, if the irreducible co-representations (co-irreps) of the occupied bands at high symmetry momenta points can be decomposed into the linear combination of MEBRs with positive integer, the system is equivalent to a magnetic AI. Otherwise, the system must be topologically nontrivial.
We have calculated the co-irreps of all occupied bands in Fig.~\ref{fig:fragile}c, which indicate that cb-AFM monolayer FeSe is a magnetic AI,
\emph{but not the AFM TI} as reported in Refs.~\cite{wang2016topological,liubin2018}.
As pointed out in Refs.~\cite{PhysRevLett.121.126402,xu2020high},
the band structures of AI have two combination of manners: i) AI = AI +AI, ii) AI =  AI + fragile topological bands.
When we focus on the one isolated flat band just blow the Fermi level,
the calculated co-irreps give rise to \{$\ols{\Gamma}_7$, $\olsi{X}_2\olsi{X}_4$, $\olsi{M}_6$\},
and its fragile topological characteristic will be discussed below.
Comparing to the MEBRs of the magnetic space group $P4'$/$n'm'm$ as tabulated in Table~\ref{table:table1},
one can find that the co-irreps of the isolated flat band can only be expressed as a combination of MEBRs with negative integer:
\begin{equation}\label{eq:fragile}
\{\ols{\Gamma}_7, \olsi{X}_2\olsi{X}_4, \olsi{M}_6\}=\olsi{E}@2c - {^2\olsi{E}_2@2a},
\end{equation}
the negative coefficient manifests that it is a fragile topological band~\cite{PhysRevLett.121.126402, PhysRevLett.120.266401}.

\begin{table}[!t]
    \caption{The MEBRs induced by maximal symmetry Wyckoff positions in $P4'$/$n'm'm$ \cite{gallego2012magnetic,xu2020high}.
    Since the monolayer FeSe is 2D, only the $z=0$ Wyckoff positions and $k_z=0$ high symmetry momenta points are considered.
    The MEBRs are denoted as \textit{D@w}, which represent a band representation induced by the \textit{D} orbital at Wyckoff position \textit{w}.}
 \label{table:table1}
 \renewcommand*{\arraystretch}{1.5}
 \centering
 \begin{tabular*}{\columnwidth}{c @{\extracolsep{\fill}}ccc}
\toprule
\hline\hline
     MEBRs & $\Gamma$(0,0) & $X$(0,$\pi$)  &  $M$($\pi$,$\pi$) \\
\hline
     ${^1\olsi{E}_1}@2a$  & $\ols{\Gamma}_7$ &  $\olsi{X}_2\olsi{X}_4$ &  $\olsi{M}_7$ \\
\hline
     ${^1\olsi{E}_2}@2a$  & $\ols{\Gamma}_6$ &  $\olsi{X}_2\olsi{X}_4$ &  $\olsi{M}_6$ \\
\hline
     ${^2\olsi{E}_1}@2a$  & $\ols{\Gamma}_7$ &  $\olsi{X}_3\olsi{X}_5$ &  $\olsi{M}_6$ \\
\hline
     ${^2\olsi{E}_2}@2a$  & $\ols{\Gamma}_6$ &  $\olsi{X}_3\olsi{X}_5$ &  $\olsi{M}_7$ \\
\hline
     $\olsi{E}@2c$      & $\ols{\Gamma}_6\oplus\ols{\Gamma}_7$ &  $\olsi{X}_2\olsi{X}_4\oplus \olsi{X}_3\olsi{X}_5$ &  $\olsi{M}_6\oplus\olsi{M}_7$ \\
\hline\hline
\bottomrule
\end{tabular*}
\end{table}

Such fragile topological characteristic can be usually revealed by the nontrivial winding of Wilson loop along some symmetric path over the whole BZ~\cite{PhysRevB.100.195135,Bouhon2020,Barry2019}.
Here, we design a series of square loops $l_i$ that satisfy $S_{4z}$ symmetry as shown by the upper right of Fig.~\ref{fig:fragile}d,
with $l_0$ at the $M$ point and $l_{end}$ corresponding to the boundary of the first BZ.
The calculated Wilson loop spectra of the fragile topological band are plotted as red lines in Fig.~\ref{fig:fragile}d,
which give rise to a conjugate pairs $\{e^{i\theta}, e^{-i\theta}\}$ for each $l_i$ yielding to the $\mathcal{PT}$ symmetry \cite{Alex2014}.
More importantly, each branch of the Wilson loop spectra evolves a $2\pi$ phase from the $M$ point to the whole BZ, leading to a nontrivial winding number 1.
We also design the other series of rectangle loops that break $S_{4z}$ symmetry as shown by the lower right of Fig.~\ref{fig:fragile}d.
The corresponding Wilson loop spectra are plotted as blue lines in Fig.~\ref{fig:fragile}d,
which obviously show that the crossing at $\theta=\pm\pi$ is opened, resulting in a trivial winding number 0.
Therefore, we conclude that the fragile topological band in cb-AFM monolayer FeSe is protected by the $S_{4z}$ symmetry,
which is first reported in this work. It may leads to different responds to disorder, comparing to the inversion symmetry protected fragile topological band~\cite{Hwang2019,song2021}.

\begin{figure}[t]
  \includegraphics[width=\columnwidth]{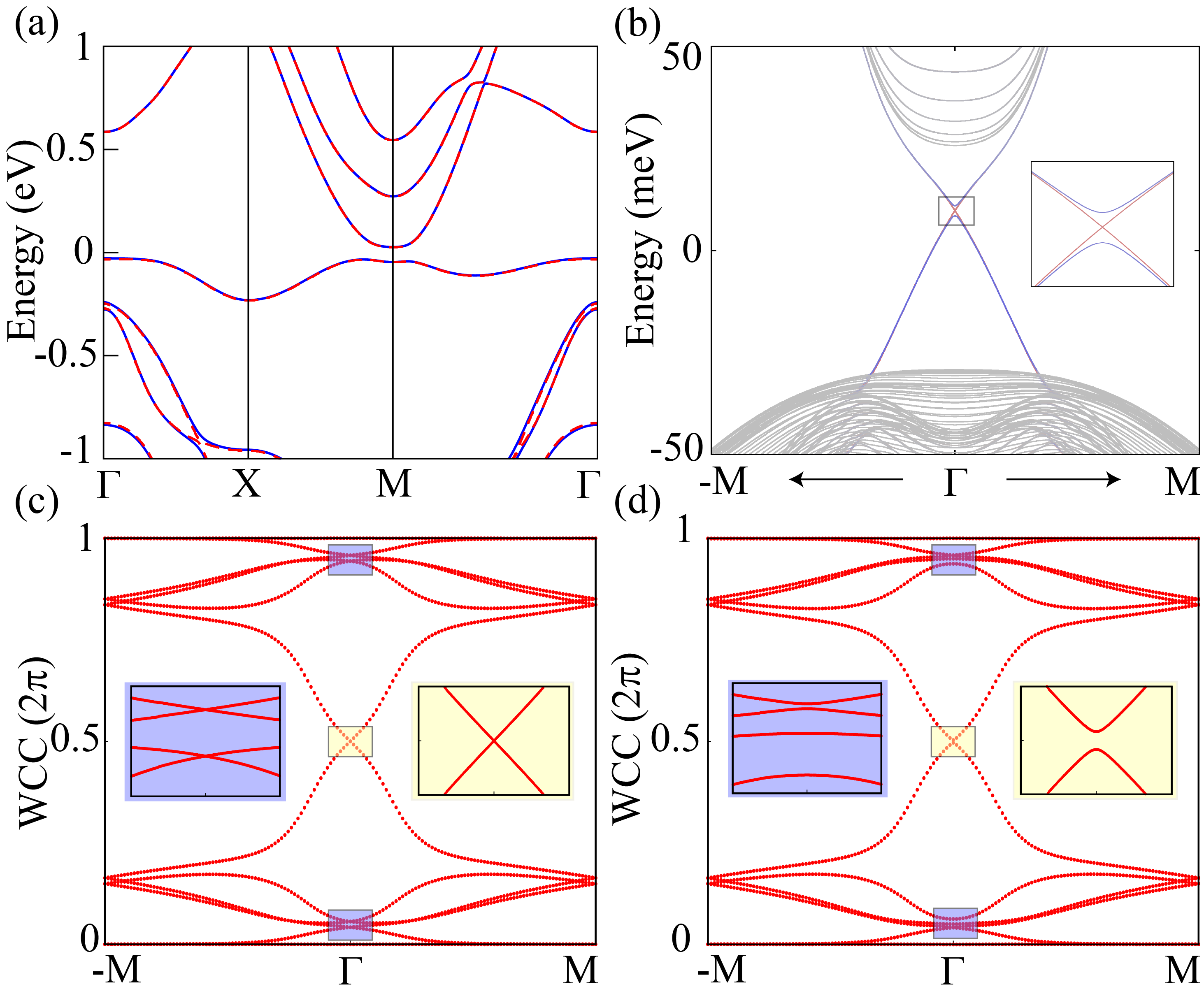}
  \caption{The electronic structures of Wannier Hamiltonian $\hat{H}_0$ and those with additional $S_z$ symmetry $\hat{H}_S$.
  (a) Band structures of $\hat{H}_S$ (red dash line) and $\hat{H}_0$ (blue solid line).
  (b) Edge states of the $\hat{H}_S$ (red) and $\hat{H}_0$ (blue), with a surface onsite energy $-0.1$~eV.
  (c)-(d) The Wilson loop spectra of all occupied bands in $\hat{H}_S$ and $\hat{H}_0$, respectively.
  }
  \label{fig:stable}
\end{figure}

In addition to the fragile topologically flat band,
we would like to point out that the cb-AFM monolayer FeSe is very close to a stable AFM TI.
Such stable AFM TI phase is protected by the spin conserving symmetry $S_z$,
which decouples the interactions between spin-up and spin-down flavor.
In general, the $S_z$ symmetry is broken in realistic electron systems~\cite{Yang2011, Li2012}.
However, we demonstrate that the $S_z$-breaking interaction in cb-AFM monolayer FeSe is very tiny,
which can be verified by two bodies of evidence.
First, we have symmetrized the Wannier Hamiltonian $\hat{H}_0$ of cb-AFM monolayer FeSe by an additional $S_z$ symmetry, i.e. $\hat{H}_S=S_z\hat{H}_0{S}_z^{-1}$,
and plot the band structures of $\hat{H}_S$ (red dash line) as a comparison to the bands of $\hat{H}_0$ (blue solid line) in Fig.~\ref{fig:stable}a.
The red bands in Fig.~\ref{fig:stable}a match the blue bands very well,
which strongly implys that the $\hat{H}_0$ is approximate to having $S_z$ symmetry.
The approximately $S_z$ symmetry is also consistent with our first-principles calculations,
which display the weak hybridization between majority spin states and minority spin states as shown in Fig.~\ref{fig:fragile}a and~\ref{fig:fragile}b.
Second, one can quantitatively estimate the $S_z$-breaking process as a competition between the SOC interaction $\lambda_{soc}$ and the on-site magnetic polarization splitting $\Delta$, where $\lambda_{soc}$ regards the spin mixing ability and $\Delta$ preserves spin conserving~\cite{Bouhon2020}.
Reminding that the magnetic splitting $\Delta$ in cb-AFM monolayer FeSe is estimated as $\sim$2.5~eV from our first-principles calculations,
and the $\lambda_{soc}$ of Fe is usually considered as $\sim$0.03~eV.
Such a huge energy difference makes that the $S_z$ symmetry is a good approximation in cb-AFM monolayer FeSe.

When $S_z$ symmetry is restored,
the system can be considered as two copies of spin-Chern insulators that are connected by the $\mathcal{PT}$ symmetry.
In Fig.~\ref{fig:stable}b, we calculate and plot the edge states of $\hat{H}_S$ on the AFM boundary (cut \#1 in Fig.~\ref{fig:crystal}c) as red lines,
which shows that two edge states crossing each other exactly and form a massless Dirac cone at the $\Gamma$ point.
As a result, we confirm that the $\hat{H}_S$ describes an AFM TI protected by the $S_z$ symmetry.
As a comparison, we also plot the calculated edge spectra of $\hat{H}_0$ on \#1 boundary as the blue lines in Fig.~\ref{fig:stable}b.
As expected, the edge spectra of $\hat{H}_0$ are very similar to those of $\hat{H}_S$,
except a very tiny massive gap about of 0.5~meV exists at the $\Gamma$ point yielding to the $S_z$ symmetry breaking.
These results confirm again that the $S_z$-breaking is very weak in the cb-AFM monolayer FeSe, which is very close to a well-defined AFM TI.
This is why the quantum spin Hall like edge states have been reported and observed in Refs.~\cite{wang2016topological,liubin2018}.
Apart from the edge states, the topological properties can also be revealed by the Wilson loop spectra of all occupied bands~\cite{PhysRevB.89.155114}.
In Fig.~\ref{fig:stable}c and~\ref{fig:stable}d, we plot the Wilson loop spectra along the $\Gamma-M$ path of $\hat{H}_S$ and $\hat{H}_0$, respectively.
The spectra in Fig.~\ref{fig:stable}c obviously show gapless crossing at the time-reversal momenta points,
reflecting a well-defined spin-Chern number $|C_s|=1$.
However, the Wilson loop spectra of $\hat{H}_0$ are generally gaped,
indicating that cb-AFM monolayer FeSe is not a well-defined AFM TI.

\textit{Corner States} ---
The fragile topology usually relates to a HOTI and leads to filling anomaly~\cite{wieder2018}.
Such filling anomaly could give rise to partially occupied corner states at the sample corner of a 2D HOTI~\cite{song2017,wieder2018}.
In the system with S$_{4z}$ symmetry, one can identify a HOTI by the S$_{4z}$ eigenvalues of occupied bands \cite{PhysRevB.100.235137, Schindlereaat0346}.
We develop a Fu-Kane-like formula to compute the HOTI indicator $Z$ as following,
\begin{eqnarray}
(-1)^Z &=& sgn(\prod_{n\in occupied}\zeta_n(\Gamma)\zeta_n(M)\zeta_n(X)),
\end{eqnarray}
where $\zeta_n(\Gamma)$ and $\zeta_n(M)$ are the $S_{4z}$ eigenvalues of the \textit{nth}-band at $\Gamma$ and $M$ point, respectively,
$\zeta_n(X)$ is the $C_{2z}$ eigenvalue of the \textit{nth}-band at $X$ point.
$Z=0$ indicates an AI, where the Wannier centers locate at Wyckoff position $2a$.
$Z=1$ indicates a HOTI, in which the Wannier centers move to $2c$ position.
The calculated results of all occupied bands are $\zeta(\Gamma)=-1$, $\zeta(M)=1$ and $\zeta(X)=1$,
which certify that cb-AFM monolayer FeSe fall into the HOTI phase with $Z=1$.

\begin{figure}[t]
  \includegraphics[width=\columnwidth]{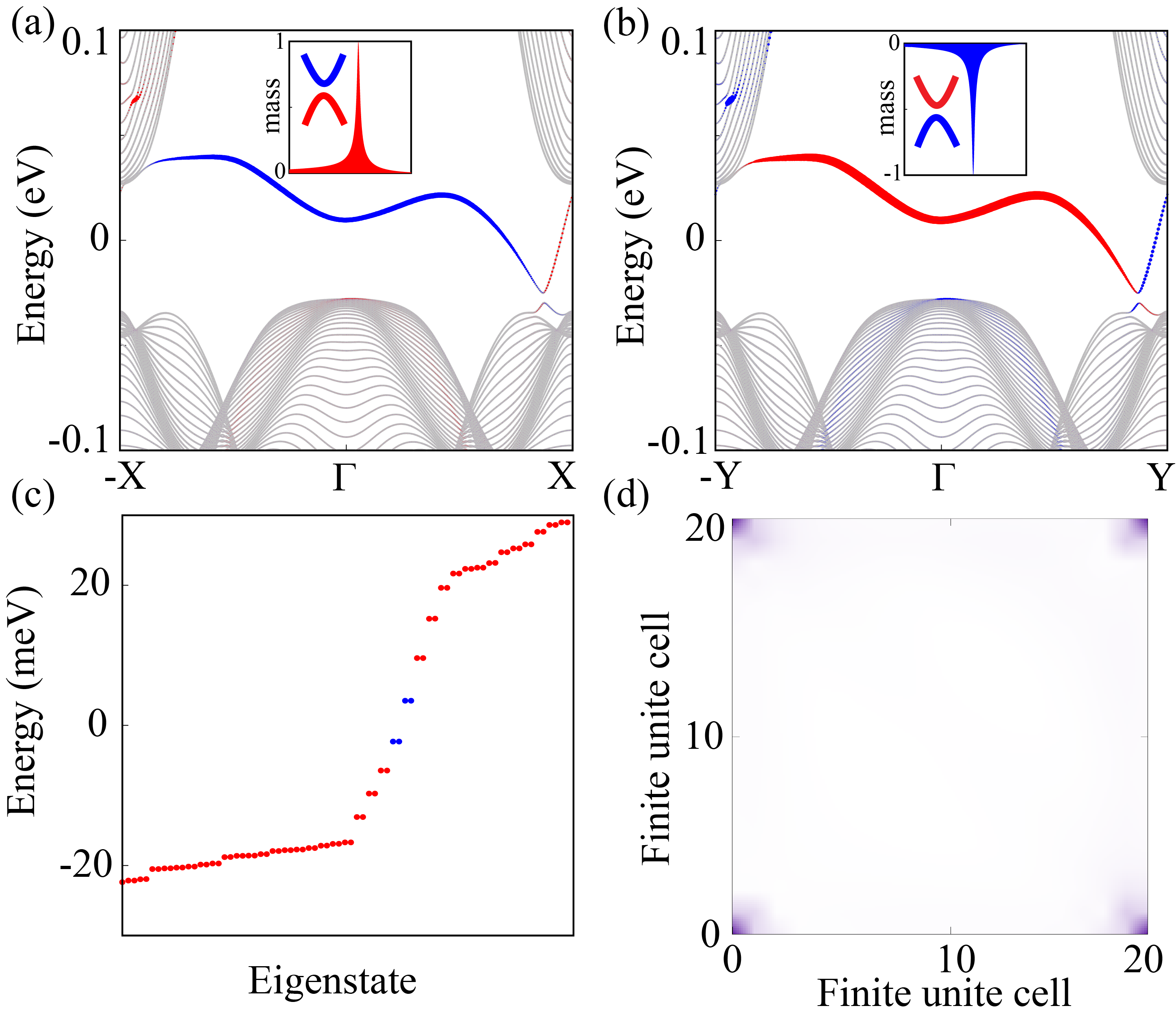}
  \caption{(a)-(b) Spin-polarized edge states along the cuts \#2 and \#3 boundary, respectively.
  The weights of positive (negative) $\sigma_z$ eigenvalue is characterized as red (blue).
  The inset pictures show the calculated mass term $\sigma_y$ near the massive Dirac cone.
  (c) Energy spectra of the $20\times20$ finite square structure. (d) Spatial distribution of the two occupied corner states.}
  \label{fig:corner}
\end{figure}

Such HOTI property can also be understood from the opposite mass term of the edge states on two different FM boundary
(cut \#2 and cut \#3 in Fig.~\ref{fig:crystal}d)~\cite{PhysRevLett.125.056402}.
We calculate and plot the spin polarized edge states along the \#2 and \#3 boundary in Fig.~\ref{fig:corner}a and~\ref{fig:corner}b, respectively.
In both of which, a massive gap can be observed between two opposite spin polarized edge states.
Further analysis shows that the gap opening on the spin-up boundary (\#2) is induced by a mass term $m_{\#2}\propto\sigma_y$,
while the gap opening on the spin-down boundary (\#3) is induced by an opposite mass term $m_{\#3}\propto -\sigma_y$,
as shown in the inset of Fig.~\ref{fig:corner}a and~\ref{fig:corner}b, respectively.
The opposite mass term is consistent with previous work that two boundaries are related by the $C_{4z}\mathcal{T}$ symmetry~\cite{Schindlereaat0346}.
Therefore, one can expect that the mass term could cancel each other at the intersection of $\#2$ and $\#3$ boundary,
and left zero-energy corner states surviving at $2c$ position.

In Fig.~\ref{fig:corner}c, we plot the low energy spectra of a square sample
with the center at $2c$ position constructed by 20~$\times$~20 unit cells by using the the Wannier Hamiltonian $\hat{H_0}$.
As expected, four corner states including two occupied and two unoccupied (blue dots in Fig.~\ref{fig:corner}c) appear near the Fermi level.
However, because $\hat{H_0}$ has no chiral symmetry, the four corner states are not pinned at zero energy exactly.
In Fig.~\ref{fig:corner}d, we plot the spatial distribution of the two occupied corner states,
which are localized at four corners and each corner has $e/2$ charge,
corresponding to previous conclusions of the HOTI protected by $C_{4z}\mathcal{T}$ symmetry very well~\cite{PhysRevB.100.235137}.
Finally, we emphasize that such calculated corner states are strongly supported by the previous experimental observations,
in which bound state near the Fermi level has been observed at the crossing point of four domain walls by scanning tunneling microscopy~\cite{liubin2018}.
We would like to suggest the experimenters to measure the charge of each bound state,
which will be very helpful to confirm its topological origin.

\textit{Discussion} ---
At last, we would like to discuss that the HOTI phase and corner states can only exist in the cb-AFM phase,
but cannot be realized in other magnetic phases such as the nonmagnetic or stripe AFM phase.
As we all know, the corner states are associated with the mismatch between the atomic sites and the charge centers~\cite{song2017}.
As shown in Fig.~\ref{fig:crystal}c and \ref{fig:crystal}d, the electrons on the nearest neighboring Fe ions are oppositely polarized in the cb-AFM configuration,
which could strongly suppress the nearest neighbor hopping $t_1$(see Fig.~\ref{fig:crystal}c),
and makes the next nearest neighbor hopping $t_2$ domination.
The dominant $t_2$ will lead to the charge centers move to the $2c$ site, and give rise to corner states at the sample corner.
However, $t_1$ is always dominant in the NM or stripe AFM configuration,
it makes the charge centers locate at the $4d$ site (see Fig.~\ref{fig:crystal}b),
where the electrons can smoothly move to $2a$ site or $2c$ site.
Therefore, our calculations, combining with the observation of the corner states~\cite{liubin2018},
strongly suggest that the cb-AFM phase is realized in the few layers of the tensile FeSe.

\textit{Acknowledgments} --- The authors thank Wei Li for valuable discussion.
This work was supported by the National Key Research and Development Program of China (2018YFA0307000), and the National Natural Science Foundation of China (11874022).

\end{document}